\documentstyle[aps,preprint,tighten,eqsecnum]{revtex}
  
\newcommand{\cn}{{\rm cn}}

\begin{document} 
  
%\input{epsf} 
  
%\draft 

\title{Semiclassical Statistical Mechanics\footnote{Invited talk at the
La Plata meeting on `Trends in Theoretical Physics', La Plata, April, 1997}}

\author{C.\ A.\ A.\ de Carvalho\footnote{e-mail: aragao@if.ufrj.br}} 

\address{Instituto de F\'\i sica, Universidade Federal do Rio de Janeiro,  
\\ Cx.\ Postal 68528, CEP 21945-970, Rio de Janeiro, RJ, Brasil} 
 
\author{R.\ M.\ Cavalcanti\footnote{e-mail: rmc@fis.puc-rio.br}} 
 
\address{Departamento de F\'\i sica, Pontif\'\i cia Universidade 
 Cat\'olica do Rio de Janeiro, \\ Cx.\ Postal 38071, CEP 22452-970, 
 Rio de Janeiro, RJ, Brasil} 
 
%\date{\today} 
 
\maketitle

\begin{abstract} 
 
We use a semiclassical approximation to derive the partition  
function for an arbitrary potential in one-dimensional  
Quantum Statistical Mechanics, which we view as an example of  
finite temperature scalar Field Theory at a point. We rely on  
Catastrophe Theory to analyze the pattern of extrema of the  
corresponding path-integral. We exhibit the propagator in the  
background of the different extrema and use it to compute the   
fluctuation determinant and to develop a (nonperturbative)  
semiclassical expansion which allows for the calculation of
correlation functions. We discuss the examples of the single  
and double-well quartic anharmonic oscillators, and the 
implications of our results for higher dimensions. 
 
\end{abstract}

\newpage

\section{Introduction} 
 
The development of nonperturbative methods for quantum field 
theories remains as much a challenge as a necessity. The many 
areas of Physics where field-theoretic approaches have proved 
useful are full of examples of phenomena whose description lies 
outside the scope of perturbation theory. In most cases, a 
reliable treatment of the nonperturbative dynamics is still 
lacking. 
 
With the exception of a limited number of exactly solvable models, 
most systems requiring nonperturbative methods have been the object 
of approximate treatments. This is an unavoidable consequence of 
the complexity of the dynamics involved. Nevertheless, it should 
not prevent us from having a clearer understanding of the physics, 
as long as we have some control over the approximations made. 
 
The approximate methods most commonly used could, perhaps, be 
classified into three major groups: expansions around special limits, 
variational methods and numerical techniques. Semiclassical and 
$1/N$ expansions, Gaussian and Hartree-Fock methods, and Monte Carlo 
techniques in lattice calculations exemplify each respective 
group. In all of them, serious technical difficulties stand in the 
way of establishing reliable controls and, in some cases, of even 
fully implementing the approximation scheme. 
 
Semiclassical expansions certainly pose hard problems for both their 
implementation and control. In this article, we discuss a physical 
situation where such problems can be overcome: one-dimensional 
Quantum Statistical Mechanics or, from a field-theoretic point of
view, scalar finite temperature Field Theory in zero spatial dimension.  
The drastic reduction in dimensionality is the price we have to pay 
to accomplish our goal: a thorough semiclassical treatment whose 
results can be compared to those of perturbation theory, as well as 
to exact ones, when available. One should not forget, however, that 
numerous applications in Physics can be studied within this framework, 
and that, as we shall indicate in the sequel, symmetries may reduce 
higher dimensional problems to situations where our results may still
be used. 
  
The article will start with a discussion of the quadratic semiclassical 
approximation (Section II\footnote{This section is strongly based on
Ref.\cite{BJP}.}), comprising an analysis of the extrema of the 
classical action and of the fluctuation determinants around them. A 
connection to Catastrophe Theory allows for the identification of the 
minima amidst the complexity of the set of classical solutions. The 
next step (Section III) is the derivation, using techniques
described in \cite{Z-J,Boyan}, of the semiclassical propagator 
in the background of any one of the extrema, a crucial result which 
leads to the expression for the fluctuation determinant and serves 
as the basis for a semiclassical expansion (Section IV). The expansion 
provides, already in first order, nonperturbative estimates for 
correlation functions. Ground states of quartic anharmonic oscillators 
can be computed to that order: the result contains a sum over an 
infinite number of diagrams in perturbation theory, an indication of 
the power of the technique. Finally (Section V), we summarize our conclusions and comment on possible applications of these 
results in higher dimensions. 

%%%%%%%%%%%%%%%%%%%%%%%%%%%%%%%%%%%%%%%%%%%%%%%%%%%%%%%%%%%%%%%%%%%%%%%%% 
 
\section{The quadratic semiclassical approximation}

\subsection{Extrema and determinants}
 
The partition function for a one-dimensional 
quantum-mechanical system in an arbitrary potential, $V(x)$, 
can be written as a path-integral\cite{Feynman-Hibbs,Feynman}:
\begin{eqnarray}  
Z(\beta)&=&\int_{-\infty}^{\infty}dx_0 
\int_{x(0)=x_0}^{x(\beta\hbar)=x_0}[Dx(\tau)]\, 
{\rm e}^{-S/\hbar}\,, \label{Z} \\  
S&=&\int_{0}^{\beta\hbar}d\tau\, 
\left[\frac{1}{2}M\dot x^2+V(x)\right]\,.  
\end{eqnarray}
As mentioned before, this can be viewed as the partition function
for a scalar finite temperature field theory, in the limiting
case where the number of space dimensions is zero.

We may expand the Euclidean action around its minima, i.e., the
stable classical trajectories,\footnote{For classical trajectories
we mean solutions of the Euler-Lagrange equation $M\ddot{x}-V'(x)=0$,
which is the equation of motion for a particle moving in the potential
{\em minus} $V$.}
and keep terms only up to
quadratic. The functional integral over the paths can, then, be
performed to yield a quadratic semiclassical evaluation of
(\ref{Z}):
\begin{equation}  
\label{Zsc} 
Z_{\rm sc}(\beta)=\int_{-\infty}^{\infty}dx_0 
\sum_{j=1}^{N(x_0,\beta)}{\rm e}^{-S_j(x_0,\beta)/\hbar} 
\Delta_j^{-1/2}(x_0,\beta)\,, 
\end{equation} 
where $S_j(x_0,\beta)$ denotes the action and $\Delta_j(x_0,\beta)$ 
represents the determinant of the fluctuation operator, 
\begin{equation} 
\label{F} 
{\hat{F}}_j=-M\frac{d^2}{d\tau^2}+V''[x_j(\tau)]\,, 
\end{equation} 
both calculated at the $j$-th classical trajectory, $x_j(\tau)$, 
satisfying the boundary conditions $x_j(0)=x_j(\beta\hbar)=x_0$; 
$N(x_0,\beta)$ is the number of classical trajectories which are 
minima of the action functional. 
  
The action has a simple expression in terms of the turning points 
of the classical trajectory: 
\begin{equation}  
\label{S}  
S_j(x_0,\beta)=\beta\hbar\,V(x_{\pm}^j)\pm  
2\int_{x_0}^{x_{\pm}^j}dx\,Mv(x,x_{\pm}^j)  
+2n\int_{x_-^j}^{x_+^j}dx\,Mv(x,x_{\pm}^j)\,.  
\end{equation} 
Here, $v(x,y)\equiv\sqrt{(2/M)[V(x)-V(y)]}$ and 
$x_+^j\,(x_-^j)$ are turning points to the right (left) of $x_0$.  
The first term in (\ref{S}) 
corresponds to the high-temperature limit of $Z(\beta)$, 
where the classical paths collapse to a point, i.e., 
$x_{\pm}^j\rightarrow x_0$. For motion in regions where the 
inverted potential is unbounded (hereafter called unbounded motion), 
$n=0$. For periodic motion, $n$ counts the number of periods and the 
second term is absent. For bounded aperiodic motion, all terms 
are present. The last two terms will be negligible for potentials 
which vary little over a thermal wavelength, 
$\lambda=\hbar\sqrt{\beta/M}$. However, by decreasing the temperature, 
quantum effects become important. 
 
For trajectories having a single turning point $(n=0)$, 
$x_{\pm}^j$ are given implicitly by 
\begin{equation} 
\label{t} 
\beta\hbar=\pm 2\int_{x_0}^{x_{\pm}^j}\frac{dx} 
{v(x,x_{\pm}^j)}\,, 
\end{equation} 
and the fluctuation determinant by 
\begin{equation} 
\label{Delta} 
\Delta_j(x_0,\beta)= 
\pm\frac{4\pi\hbar^2[V(x_{\pm}^j)-V(x_0)]} 
{MV'(x_{\pm}^j)}\,\left[ 
\frac{\partial\beta}{\partial x_{\pm}^j}\right]_{x_0}\,. 
\end{equation} 
For trajectories with two turning points, we may generalize
(\ref{Delta}) to:
\begin{equation} 
\label{Delta2} 
\Delta_j(x_0,\beta)= 
\frac{4\pi\hbar^2[V(x_{\pm}^j)-V(x_0)]} 
{M}\,\left\{\frac{n_{+}}{V'(x_{+}^j)} \left[ 
\frac{\partial\beta}{\partial x_{+}^j}\right]_{x_0}
-\frac{n_{-}}{V'(x_{-}^j)} \left[ 
\frac{\partial\beta}{\partial x_{-}^j}\right]_{x_0}
\right\}\,.
\end{equation} 
$n_{\pm}$ counts the number of visits to the two turning points,
$x_{\pm}^j$. As we shall see in the sequel, trajectories
with more than one turning point are naturally excluded from the
calculation. The derivations of (\ref{Delta}) and (\ref{Delta2})
will be outlined in section III.

Formulae (\ref{S}), (\ref{Delta}) and (\ref{Delta2}) do not
require full knowledge of the classical trajectory, as the dependence 
on $x_j(\tau)$ comes only through the turning point. Furthermore,
equations (\ref{Delta}) and (\ref{Delta2}) yield a direct evaluation
of the determinant which bypasses solving an eigenvalue problem for
each stable classical trajectory. To see how this can 
simplify the evaluation of $Z_{\rm sc}(\beta)$, let us take 
the harmonic oscillator, $V(x)=\frac{1}{2}M\omega^2 x^2$, as an 
example. In this case, given $x_0$ and $\beta$ there is only one 
trajectory, with a single turning point, given by 
$x_+(x_-)=x_0/\cosh(\beta\hbar\omega/2)$ for $x_0<0\;(>0)$. 
$S$ and $\Delta$ 
can also be readily calculated, the final result being 
\begin{equation} 
\label{ZHO} 
Z_{\rm sc}(\beta)=\int_{-\infty}^{\infty}dx_0\, 
{\rm e}^{-(M\omega x_0^2/\hbar)\tanh(\beta\hbar\omega/2)}\, 
\sqrt{\frac{M\omega}{2\pi\hbar\sinh(\beta\hbar\omega)}}\;, 
\end{equation}
which in this case is exact.
 
For the single-well quartic anharmonic oscillator, 
$V(x)=\frac{1}{2}M\omega^2 x^2 +\frac{\lambda}{4} x^4$, $\lambda>0$, 
it is still true 
that there is only one trajectory, given $x_0$ and $\beta$, which has a 
single turning point. Eq.\ (\ref{t}) yields: 
\begin{eqnarray} 
x_{\pm}&=&x_0\,\cn(u_{\pm},k_{\pm}), \\ 
u_{\pm}&=&\frac{\beta\hbar\omega}{2}\,\sqrt{1+ 
\frac{\lambda x_{\pm}^2}{M\omega^2}}, \\ 
k_{\pm}^2&=&\frac{M\omega^2 +(1/2)\lambda x_{\pm}^2} 
{M\omega^2 + \lambda x_{\pm}^2}, 
\end{eqnarray} 
where $\cn(u,k)$ is one of the Jacobian elliptic functions\cite{GR}.  
We may, then, use Eqs.\ (\ref{S}) and (\ref{Delta}), and change 
the integration variable in (\ref{Zsc}) from $x_0$ to $x_{\pm}$ 
to obtain $Z_{\rm sc}(\beta)$. The resulting integral can be evaluated 
numerically.

A more interesting situation occurs in the case 
of the double-well quartic anharmonic 
oscillator, $V(x)=\lambda(x^2-a^2)^2$. For $x^2>a^2$, only 
single paths with single turning points exist for fixed $x_0$ 
and $\beta$. However, there is also a region, 
$x^2<a^2$, where the classical motion is bounded and a much richer 
structure exists\cite{Anker}, one in which more 
than one classical path may exist for given 
values of $x_0$ and $\beta$. 
 
In a region of bounded classical motion 
(a well in $-V$), such as $x^2<a^2$ for the anharmonic oscillator, 
the number of classical trajectories changes as the temperature drops. 
If $0\le\beta<\pi/\hbar\omega_m$ (where 
$\omega_m\equiv\sqrt{-V''(x_m)/M}$ and $x_m$ is a local minimum 
of $-V$), for every $x_0$ in this region 
there is only one closed path, with a single turning point, 
satisfying the classical equations of motion. 
For $x_0<x_m\;(>x_m)$ 
this path goes to the left (right) and returns to $x_0$. 
For $x_0=x_m$, it sits still 
at the bottom of the well. It is this single-path 
regime which goes smoothly into the high-temperature limit. 
 
For $\beta=\pi/\hbar\omega_m$, the solution that sits still 
at $x_m$ becomes unstable. Its fluctuation operator is that of 
a harmonic oscillator with $\omega^2=\omega_m^2$. Its finite 
temperature determinant is $\Delta(x_m,\beta)= 
2\pi\hbar\sin(\beta\hbar\omega_m)/M\omega_m$. 
This goes through zero at $\beta=\pi/\hbar\omega_m$ and becomes negative for 
$\beta>\pi/\hbar\omega_m$, thus signaling that 
$x(\tau)=x_m$ becomes unstable. At the same time, two new classical paths 
appear, symmetric with respect to $x_m$, as depicted in Fig.~1. 
Therefore, at $x_0=x_m$, we go from a 
single-path regime to a triple-path regime as we cross 
$\beta=\pi/\hbar\omega_m$. The two new paths are degenerate minima, whereas 
the path that sits still at $x_m$ becomes a saddle-point of the 
action, with a single negative mode. 
 
As $\beta$ grows beyond $\pi/\hbar\omega_m$, an analogous situation 
occurs for other values of $x_0$ inside the well. When the fluctuation 
determinant around the classical path for a given $x_0$ vanishes, 
a new classical path appears. Its single turning point lies opposite, 
with respect to $x_m$, to that of the formerly unique path.
If $\beta$ is increased further, this 
trajectory splits into two, as illustrated in Fig.~2. 
One is a local minimum of the action, whereas the other is a 
saddle-point, with only one negative mode. Again, we have 
transitioned from a single to a triple-path regime. As $\beta$ grows, 
the triple-path region spreads out around 
$x_m$. The frontiers of that region are defined by the 
points $x_0$ such that 
$\left({\partial\beta}/{\partial x_{\pm}}\right)_{x_0}=0$, where
the expression (\ref{Delta}) for the determinant vanishes. 
 
The phenomenon just described is an example of catastrophe.  
It takes place whenever the number of classical trajectories 
changes, with one or more of them becoming unstable, as we lower 
the temperature. Conversely, we may say that the phenomenon is 
characterized by the coalescence of two or more of the classical 
trajectories, as we increase the temperature. This is akin to 
the ocurrence of caustics in Optics\cite{Berry}, where light 
rays play the role of classical trajectories and the action 
is replaced with the optical distance. In the next subsection,
we shall use the mathematics of catastrophes to understand how
the set of classical trajectories changes with temperature.

\subsection{Catastrophe Theory} 
 
If we use the basis of eigenfunctions 
of the fluctuation operator around classical paths, and
denote by $s_1$ the coordinate associated with the direction 
of instability of our example, the action can be viewed as 
$S=S(s_1,\ldots;x_0,\beta)$, with the dots referring to all others.   
Catastrophe Theory\cite{Berry,Saunders} allows us 
to write down the ``normal form'', $S_N$, 
of this generating function in the three-dimensional subspace 
made up by the unstable direction of state 
space (the set of paths) and the two control variables, $\beta$ and $x_0$; 
it is this subspace which 
is relevant for the study of the onset of instability. As catastrophes 
are classified by their codimension in control space, which here is 
two-dimensional, only those of codimension 1 (the fold) or 2 (the cusp) 
are generic (i.e., structurally stable). 
The pattern of extrema then leads to the cusp, whose generating function is 
\begin{equation} 
\label{SN} 
S_N(s_1;u_1,v_1)=s_1^4+u_1s_1^2+v_1s_1\,, 
\end{equation} 
where $u_1$ and $v_1$, the control parameters, are related to 
$\beta$ and $x_0$, respectively. This catastrophe is defined by 
\begin{equation} 
\label{SN'} 
\frac{\partial S_N}{\partial s_1}= 
\frac{\partial^2S_N}{\partial s_1^2}=0\,. 
\end{equation} 
Eliminating $s_1$ from these equations, we can draw the bifurcation 
set in control space (the curve $v_1^2=-4u_1^3/27$), as well as 
the pattern of extrema of $S_N$ (Fig.~3). Assuming that $u_1$ and 
$v_1$ are linearly related to $(\beta-\pi/\hbar\omega_m)$ and 
$(x_0-x_m)$, respectively, for small values of both, we can 
also schematically plot the classical action (i.e., the action for 
classical trajectories) as a function of $x_0$ for different values of 
$\beta$;\footnote{This plot can be obtained by exploiting the 
relation between the cusp and 
the swallowtail catastrophes. The latter has a generating function 
whose normal form is $W(s_1;a,b,c)=s_1^5+as_1^3+bs_1^2+cs_1$, where 
$a$, $b$, $c$ are control variables. The extremum condition 
is, then, ${\partial W}/{\partial s_1}=0$. 
 The identifications $S_N\equiv -c/5$, $u_1\equiv 3a/5$ and 
$v_1\equiv 2b/5$ will make the additional condition that defines 
the swallowtail, $\partial^2W/\partial s_1^2=0$, coincide with 
the requirement $\partial S_N/\partial s_1=0$. (In the jargon of 
catastrophe theory: the bifurcation set for the swallowtail coincides 
with the equilibrium hypersurface of the cusp.)} 
see Fig.~4(a,b). 
 
New classical trajectories, accompanied by new 
catastrophes, will occur as we keep increasing $\beta$. From 
$\beta=2\pi/\hbar\omega_m$, we start having trajectories 
with $\dot x(0)=0$, i.e., with one full period. For these, the 
determinant of fluctuations vanishes because the first, not the 
second bracket of (\ref{Delta}) goes trough zero. However, the fluctuation 
operator around such trajectories already has a negative 
eigenmode. This follows from the fact that the zero-mode, 
given by $\dot x_{\rm cl}(\tau)$, has a node. Thus, it 
cannot be the ground-state for the associated Schr\"odinger 
problem. This means that a new catastrophe 
takes place along a second direction in function space. 
 
This new catastrophe is also a cusp. At $x_m$, 
as $\beta$ becomes larger than $2\pi/\hbar\omega_m$, 
the solution that sits still acquires a second negative eigenvalue 
--- thus becoming unstable along a new direction in function space --- 
and two other solutions appear. They both have a negative 
eigenvalue along the first direction of instability 
and a positive eigenvalue along the new direction of instability.  
If we combine the information from both directions, 
we find that we go from two minima and a saddle-point with only 
one negative eigenvalue (hereafter, a one-saddle) to two minima, 
two one-saddles and one two-saddle. The two one-saddles are 
degenerate in action and represent time-reversed periodic paths. As 
$\beta$ increases beyond $2\pi/\hbar\omega_m$, the same phenomenon 
takes place for $x_0$ around $x_m$: the one-saddles that already 
existed for $\beta<2\pi/\hbar\omega_m$ in the three-path 
region around $x_m$ become unstable along a second direction, 
and a five-path region grows around $x_m$, with two minima, 
two (periodic, degenerate in action) one-saddles and a 
two-saddle. 
 
The new cusp can be cast into normal form using the second direction of 
instability: 
\begin{equation} 
\label{SN2} 
S_N(s_2;u_2,v_2=0)=s_2^4+u_2s_2^2\sim(s_2^2+\frac{u_2}{2})^2\,. 
\end{equation} 
The absence of the linear term in Eq.\ (\ref{SN2}) [compare with 
Eq.\ (\ref{SN})] reflects the degeneracy in action of the two 
one-saddles. However, there is another degeneracy: since $V$ does 
not depend on $\beta$, if $x_{\rm cl}(\tau)$ is a solution of the 
Euler-Lagrange equation, so is $x_{\rm cl}(\tau+\tau_0)$. If 
$x_{\rm cl}(\tau)$ is periodic, $x_{\rm cl}(\tau+\tau_0)$ 
describes the same path (and so has the same action), but with 
another starting point. 
This can be represented in Eq.\ (\ref{SN2}) by 
the choice $u_2\propto [(x_0-x_m)^2+k(2\pi/\hbar\omega_m-\beta)]$, 
with $k>0$. See Fig.\ 4(c) 
 
As we approach $\beta=3\pi/\hbar\omega_m$, the situation becomes 
similar to the one near $\beta=\pi/\hbar\omega_m$. The difference is 
that we now have to deal with a path with more 
than one and less than two periods. Near $\beta=4\pi/\hbar\omega_m$, 
two-period paths intervene, a situation similar to that at 
$\beta=2\pi/\hbar\omega_m$. The pattern which develops\cite{Peixoto} 
is depicted in Fig.~5. 
 
Despite the fact that we keep adding new extrema as we lower the 
temperature, only 
two of these are stable (i.e., minima). In a semiclassical 
approximation in euclidean time, appropriate to equilibrium 
situations, these are the only extrema we 
have to sum over, meaning that $N(x_0,\beta)$ never exceeds two.  
For $\beta>\pi/\hbar\omega_m$, there will be regions with either one 
or two minima of the classical action. The transition from a 
single-minimum to a double-minimum region occurs at values of $x_0$ 
where the fluctuation determinant vanishes due to the appearance of 
a caustic, thus leading to a singularity in 
the integrand of Eq.\ (\ref{Zsc}). This is not a disaster, 
however, as this singularity is integrable. 
(Such a singularity is an artifact of the semiclassical approximation. 
It disappears in a more refined approximation\cite{Berry,DV,Schulman,Weiss}, 
in which one includes higher order fluctuations in the direction(s) of 
function space where the instability sets in.) We shall exhibit the 
details of this calculation in a forthcoming publication\cite{joras}. 
 
To conclude, two remarks: (i) our results may be used to derive a 
semiclassically improved perturbation expansion, which will be 
presented in Section IV; (ii) the semiclassical partition function 
incorporates, in an approximate way, quantum effects that become 
more and more relevant with the increase of the thermal wavelength.  
At very low temperatures, however, the quadratic approximations 
inherent to the semiclassical approximation fail to capture 
the details of the potential, which are important in the regime 
of large thermal wavelengths. Thus, we expect our calculations 
to be a better approximation at high temperatures. Indeed, studying 
the $T=0$ limit of $Z_{\rm sc}(\beta)$ for the anharmonic oscillator, 
one finds no corrections to the unperturbed ground state energy.  
Nevertheless, this can be corrected by using the semiclassically 
improved perturbation theory just mentioned. 

%%%%%%%%%%%%%%%%%%%%%%%%%%%%%%%%%%%%%%%%%%%%%%%%%%%%%%%%%%%%%%%%%%%%%%%
 
\section{The Semiclassical Propagator} 
 
In the next section we will show that, in order to go beyond the quadratic 
approximation of the preceding section, it is necessary to derive the semiclassical propagator, defined as: 
\begin{eqnarray} 
& &G_j(\eta_1,\eta_2;\tau_1,\tau_2)\equiv
\int_{\eta(\tau_1)=\eta_1}^{\eta(\tau_2)=\eta_2} 
[D\eta(\tau)]\,e^{-\Sigma_j/\hbar}\;, 
\label{Gj} \\
& &\Sigma_j\equiv\frac{1}{2}\int_{\tau_1}^{\tau_2}d\tau\,\left\{M{\left(\frac
{d\eta} {d\tau}\right)}^2+V''[x_j(\tau)]\,\eta^2\right\}\;. 
\end{eqnarray}
$\Sigma_j$ involves the fluctuation operator, $\hat{F}_j$, which corresponds
to the second functional derivative of the action around the classical 
trajectory, $x_j(\tau)$. 
 
Formula  (\ref{Gj}) is the path-integral expression for the 
Euclidean time evolution operator, $\hat{\rho}_j$, of the quantum-mechanical 
problem defined by 
\begin{eqnarray} 
& &-\hbar\frac{\partial}{\partial\tau}\,\hat{\rho}_j(\tau,0)=\hat{H}_j(\tau)
\,\hat{\rho}_j(\tau,0)\;, 
\label{Hb} \\
& &\hat{H}_j(\tau)\equiv -\frac{\hbar^2}{2M}\frac{\partial^2}{\partial\eta^2}+
\frac{1}{2} 
V''[x_j(\tau)]\,\eta^2\;. 
\label{H} 
\end{eqnarray}
Indeed, the propagator is given by 
\begin{equation} 
G_j(\eta_1,\eta_2;\tau_1,\tau_2)=\langle \eta_2| \hat{\rho}_j(\tau_2,\tau_1)|\eta_1\rangle\;. 
\label{prop1} 
\end{equation} 
Thus, it is just a density matrix element, as suggested by equation (\ref {H}). 
 
As the integral in  (\ref{Gj}) is Gaussian ($\hat{H}_j$ is quadratic), it is
completely determined in terms of the extremum, $\bar{\eta}_j(\tau)$, of
$\Sigma_j[\eta(\tau)]$, which satisfies
\begin{equation} 
-M\,\frac{d^2\bar{\eta}_j}{d\tau^2}+V''[x_j(\tau)]\,\bar{\eta}_j=0\,,
\label{gauss} 
\end{equation}
with the boundary conditions  
\begin{equation}
\bar{\eta}_j(\tau_1)=\eta_1\quad {\rm and}\quad\bar{\eta}_j(\tau_2)=\eta_2\,. 
\label{bc}
\end{equation}
The semiclassical propagator is, then,
\begin{equation} 
G_j(\eta_1,\eta_2;\tau_1,\tau_2)=G_j(0,0;\tau_1,\tau_2)\,
e^{-\bar{\Sigma}_j/\hbar}\;, 
\label{prop2} 
\end{equation} 
with $\bar{\Sigma}_j\equiv\Sigma_j[\bar{\eta}_j(\tau)]$. Using equations 
(\ref{Gj}), (\ref{gauss}) and (\ref{bc}), we have
\begin{equation} 
\bar{\Sigma}_j=\frac{M}{2}\left[\eta_2 \dot{\bar{\eta}}_j(\tau_2) - 
\eta_1 \dot{\bar{\eta}}_j(\tau_1)\right]. 
\end{equation} 
The fluctuation determinant discussed in the previous section can be 
obtained as
\begin{equation} 
\Delta_j^{-1/2}(x_0,\beta)=G_j(0,0;0,\beta\hbar)\;. 
\end{equation} 
 
The extremum, $\bar{\eta}_j(\tau)$, will be the linear combination of two 
linearly  independent solutions of equation (\ref {gauss}), $a_j(\tau)$ and 
$b_j(\tau)$, which satisfies the boundary conditions (\ref{bc}):
\begin{equation} 
\bar{\eta}_j(\tau)=\frac{\eta_1 b_{j2}-\eta_2 b_{j1}}
{a_{j1}b_{j2}-a_{j2}b_{j1}}\,a_j(\tau) + 
\frac{\eta_2 a_{j1}-\eta_1 a_{j2}}{a_{j1}b_{j2}-a_{j2}b_{j1}}\,b_j(\tau)\;, 
\end{equation} 
with $a_{ji}\equiv a_j(\tau_i)$ and $b_{ji}\equiv b_j(\tau_i)$. This allows us 
to compute  $\bar{\Sigma}_j$. Introducing the functions
\begin{eqnarray} 
{\cal F}_{j1}(\tau)&\equiv& a_{j1}b_{j}(\tau)-a_j(\tau)b_{j1}\,, \\
{\cal F}_{j2}(\tau)&\equiv& a_{j2}b_{j}(\tau)-a_j(\tau)b_{j2}\,,
\end{eqnarray} 
which satisfy ${\cal F}_{j1}(\tau_1)=0$, ${\cal F}_{j2}(\tau_2)=0$ and
${\cal F}_{j1}(\tau_2)=- {\cal F}_{j2}(\tau_1)$, and the Wronskian
\begin{equation} 
W_j(\tau)\equiv a_j(\tau)\dot b_j(\tau)-\dot a_j(\tau)b_j(\tau)\,, 
\end{equation} 
we may write
\begin{equation}
\bar{\Sigma}_j=\frac{M}{2}\,\sum_{a,b=1}^{2}s_{ab}^{(j)}\eta_a\eta_b\,, 
\end{equation} 
where
\begin{equation} 
s_{11}^{(j)}\equiv -\frac{\dot{\cal F}_{j2}(\tau_1)}
{{\cal  F}_{j2}(\tau_1)}\,,\quad
s_{12}^{(j)}\equiv \frac{W_j(\tau_1)}{{\cal  F}_{j2}(\tau_1)}\,,\quad
s_{21}^{(j)}\equiv -\frac{W_j(\tau_2)}{{\cal  F}_{j1}(\tau_2)}\,,\quad
s_{22}^{(j)}\equiv \frac{\dot{\cal F}_{j1}(\tau_2)}{{\cal  F}_{j1}(\tau_2)}\,.
\end{equation} 
Using equations (\ref{H}), (\ref{Hb}), (\ref{prop1}) and (\ref{prop2}), 
we obtain
\begin{equation} 
\left[\frac{\partial}{\partial\tau_2}+\frac{1}{2}\,\frac{\dot{\cal F}_{j1}
(\tau_2)} 
{{\cal F}_{j1}(\tau_2)}\right]G_j(0,0;\tau_1,\tau_2)=0\,, 
\end{equation} 
which leads to
\begin{equation} 
G_j(0,0;\tau_1,\tau_2)=C_j(\tau_1)\,[{\cal  F}_{j1}(\tau_2)]^{-1/2}. 
\label{GF} 
\end{equation} 
As $\tau_2\to\tau_1$, we must recover the free-particle result: 
\begin{equation} 
\lim_{\tau_2\to\tau_1}G_j(0,0;\tau_1,\tau_2)=
{\left[\frac{M}{2\pi\hbar(\tau_2-\tau_1)}\right]}^{1/2} 
\label{limit} 
\end{equation} 
Expanding ${\cal F}_{j1}(\tau_2)$ in equation (\ref{GF}) around $\tau_1$,  
and using equation  
(\ref{limit}), we can determine $C_j(\tau_1)$. Finally:
\begin{equation} 
G_j(0,0;\tau_1,\tau_2)=\left[\frac{M}{2\pi\hbar}\, 
\frac{\dot{\cal F}_{j1}(\tau_1)}{{\cal  F}_{j1}(\tau_2)}\right]^{1/2}. 
\end{equation} 
All that remains is to find the two linearly independent solutions of (\ref{gauss}),  
$a_j(\tau)$ and $b_j(\tau)$. The arguments of section II allow us to restrict
our attention to single turning point paths, although we will sketch how one 
should proceed in the most general case, at the end of this section. 
 
It can be easily checked that $\dot x_j(\tau)$ satisfies 
equation (\ref {gauss}), 
by simply differentiating the Euclidean classical equation of motion with 
respect to Euclidean time. Then: 
\begin{equation} 
a_j(\tau)=\dot x_j(\tau). 
\end{equation} 
If we consider $\tau\leq\beta\hbar/2$, one can check that
\begin{equation} 
c_j\equiv\dot x_j(\tau)\int_0^\tau\frac{dt}{{\dot x_j}^2(t)} 
\end{equation} 
also satisfies equation (\ref{gauss}), again thanks to the equation of motion 
satisfied by $\dot x_j(\tau)$. Therefore, it can be used as $b_j(\tau)$ for 
$\tau\leq\beta\hbar/2$. It can be shown that $c^j_\pm\equiv c_j(\beta\hbar/2)$ 
and $\dot c^j_\pm\equiv \dot c_j(\beta\hbar/2)$ are well-defined\cite{joras}. 
%The former is
% 
%\begin{equation} 
%c^j_\pm\equiv c_j(\beta\hbar/2)=-\frac{M}{V'(x^j_\pm)}\,, 
%\end{equation} 
%
%whereas the latter is obtained by taking a careful limit from below 
%($\tau \to\beta\hbar/2-0$):
%\begin{equation} 
%\dot c^j_\pm\equiv \dot c_j(\beta\hbar/2)=\lim_{x\to x_\pm^j}\sqrt{\frac{2}{M} 
%[V(x)-V(x_\pm^j)]}\int_{x_0}^{x}\frac{dy}{\left\{
%\frac{2}{M}[V(x)-V(x_\pm^j)]\right\}^{3/2}}\,.
%\end{equation} 
%
Clearly, these will be the values of $b_j$ and $\dot b_j$ at $\beta\hbar/2$. 
In order to find our other solution, for $\tau>\beta\hbar/2$, we cannot use 
$c_j(\tau)$, which is ill-defined beyond $\beta\hbar/2$. However, we may use
\begin{equation} 
\tilde{c}_j(\tau)\equiv\dot x_j(\tau)\int _{\beta\hbar}^{\tau}\frac{dt}{\dot 
x_j^2(t)}\,. 
\end{equation} 
A linear combination of $a_j(\tau)$ and $\tilde{c}_j(\tau)$ may, then, be
constructed whose value at $\beta\hbar/2$ and that of its derivative coincide 
with $b_j(\beta\hbar/2)\equiv c_j(\beta\hbar/2)$ and $\dot b_j(\beta\hbar/2) 
\equiv \dot c_j(\beta\hbar/2)$. These two conditions determine the linear 
combination and guarantee continuity. As a  result, we may now define $b_j(\tau)$ 
for all values of $\tau$: 
\begin{equation}
b_j(\tau)=\cases{
\dot{x}_j(\tau)\int_0^\tau\frac{dt}{\dot{x}_j^2(t)} & for 
$\tau\leq\beta\hbar/2$, \cr 
\dot x_j(\tau)\int_{\beta\hbar}^\tau \frac{dt}{\dot x_j^2(t)}-2c_\pm^j 
\dot c_\pm^j\dot x _j(\tau) & for
$\tau \geq \beta\hbar/2$.}
\end{equation}
We note that both $a_j(\tau)$ and $b_j(\tau)$ can be completely expressed in terms of the potential and of the value of the (single) turning point. 
Also, for the 
general case of two turning points, we can use the above procedure to obtain a 
piecewise construction of  $b_j(\tau)$. For the first piece, we trade 
$\beta\hbar/2$ for the 
$\tau^\star$ where the first turning point is visited. Then, $\beta\hbar$ will 
be traded for $2\tau^\star$. This will allow us to define $b_j(\tau)$ up to $2\tau^\star$. 
We may proceed in the same fashion to the next turning point and repeat the 
operation to account for the number of cycles involved.

Equation (\ref{Delta2}) was obtained from $G_j(0,0;0,\beta\hbar)$, for the 
general case. From the knowledge of $\bar\eta_j(\tau)$, we can write a final 
expression for $G_j(\eta_1,\eta_2;\tau_1,\tau_2)$:
\begin{eqnarray}
G_j(\eta_1,\eta_2;\tau_1,\tau_2)&=&\frac{1}{\sqrt{2\pi\sigma_j^2}}\,
e^{-\frac{1}{2\sigma_j^2}\left[ A_{12}^j\eta_2^2+A_{21}^j\eta_1^2-
2\eta_1\eta_2\right]} \\
\sigma_j&=&\frac{\hbar}{M}\,{\cal F}_{j1}(\tau_2)\,, \\
A_{12}^j&=&\frac{a_j(\tau_1)+\dot a_j(\tau_2)
{\cal F}_{j1}(\tau_2)}{a_j(\tau_2)}\,.
\end{eqnarray}
This will be the basis for the semiclassical expansion to be derived
in our next section.
 
%%%%%%%%%%%%%%%%%%%%%%%%%%%%%%%%%%%%%%%%%%%%%%%%%%%%%%%%%%%%%%%%%%%%%%

\section{The Semiclassical Expansion}

We are now in a position to go beyond the quadratic approximation. To do
so, we expand the action around a classical trajectory, $x_j(\tau)$, to 
all orders. Letting $x(\tau)=x_j(\tau)+\eta(\tau)$, with 
$\eta(0)=\eta(\beta\hbar)=0$, we obtain:
\begin{equation}
S[x(\tau)]=S[x_j(\tau)]+\frac{1}{2!}\int_0^{\beta\hbar} d\tau \, \eta(\tau)\,
\hat{F}_j\,\eta(\tau) + \sum_{n=3}^{\infty}\frac{1}{n!}\int_0^{\beta\hbar}
d\tau \, V^{(n)}[x_j(\tau)]\,\eta^n(\tau),
\end{equation}
where $V^{(n)}\equiv d^nV/dx^n$, and the second term is the quadratic 
piece. Use was made of the fact that higher functional derivatives 
involve products of delta functions.

As we take $e^{-S/\hbar}$ and integrate over the deviations, $\eta(\tau)$, 
from the classical trajectory, we may expand the higher than quadratic 
terms in a power series. Defining:
\begin{equation}
\int_0^{\beta\hbar}d\tau \, \delta L_j\equiv \sum_{n=3}^{\infty}\frac{1}{n!}
\int_0^{\beta\hbar}d\tau \,  V^{(n)}[x_j(\tau)]\,\eta^n(\tau),
\end{equation}
we may interchange the integral over $\tau$ above with the sums over 
deviations from the classical path. This has the effect of 
breaking up the path integral 
into pieces that go from zero to the various times corresponding to 
multiple insertions of $\delta L_j$ and, finally, to $\beta\hbar$. The 
result of doing the path integral in each piece yields the semiclassical 
propagator at different time intervals. The expression for the partition 
function becomes:
\begin{eqnarray}
\label{particao4}
Z(\beta)&=&\int_{-\infty}^{\infty}dx_0\sum_{j=1}^{N(x_0,\beta)}
e^{-S_j/\hbar}\,[ G_j(0,0;0,\beta\hbar)
\nonumber \\
& &- \frac{1}{\hbar} 
\int_0^{\beta\hbar}d\tau_1\,\int_{-\infty}^{\infty}d\eta_1 \,
G_j(0,\eta_1;0,\tau_1)\,\delta L_j(\tau_1,\eta_1)\,
G_j(\eta_1,0;\tau_1,\beta\hbar)
\nonumber \\
& &+  \frac{1}{2\hbar^2} \int_0^{\beta\hbar}d\tau_2\,
\int_{-\infty}^{\infty}d\eta_2\,
\int_0^{\tau_2}d\tau_1\,\int_{-\infty}^{\infty}d\eta_1\,
G_j(0,\eta_1;0,\tau_1)\,\delta L_j(\tau_1,\eta_1) 
\nonumber \\
& &\times G_j(\eta_1,\eta_2;\tau_1,\tau_2)\,\delta L_j(\tau_2,\eta_2)\,
G_j(\eta_2,0;\tau_2,\beta\hbar)] + O(3)
\end{eqnarray}
Clearly, the Gaussian dependence of the semiclassical propagator on the 
$\eta$ variables, together with the fact that $\delta L_j$ only involves 
powers of $\eta$, allows the integrals over $\eta$ to be easily performed. 
As for the $\tau$ dependence, it is completely known, although the 
integrals involved may not be done analytically. The integrals over $\tau$ 
can, alternatively, be transformed into integrals over $x$ (the classical 
position). We should note that the first term in 
(\ref{particao4}) is just the fluctuation determinant and that each 
individual term in the series corresponds to an infinite sum of diagrams 
in ordinary perturbation theory, plus additional terms which depend on
the classical solution.

In order to illustrate the usefulness of the expansion, let us outline 
some of its consequences in the case of the single-well quartic anharmonic 
oscillator, $V(x)=\frac{1}{2}m\omega^2x^2+\frac{\lambda}{4}x^4$, already 
discussed in section II. If we define a dimensionless coupling $g\equiv 
\lambda\hbar/2m^2\omega^3$, the first order correction in the expansion, 
i.e., the second term in the bracket of equation (\ref{particao4}), may be 
written as $-3g\,G_j(0,0;0,\beta\hbar)\,I_j(\beta,g)$,
where $I_j$ is a dimensionless integral:
\begin{equation}
I_j\equiv \frac{\omega^3}{(2c_{\pm}^j\dot c_{\pm}^j)^2}
\int_0^{\beta\hbar} d\tau \, \left[ b_j^4(\tau)+4c_{\pm}^j\dot c_{\pm}^j
b_j^3(\tau)a_j(\tau)+(2c_{\pm}^j\dot c_{\pm}^j)^2b_j^2(\tau)a_j^2(\tau)
\right]
\end{equation}
The functions $a_j(\tau)$ and $b_j(\tau)$ can be expressed in terms of 
Jacobian elliptic functions and elliptic integrals \cite{joras} which 
we shall not write down. All we want to emphasize is that $I_j(\beta,g)$ 
can be expanded in a power series in $g$, just as $G_j(0,0;0,\beta\hbar)$. 
In ordinary perturbation theory, the first order term is of order $g$, 
and goes at most as $\beta$, for large $\beta$. An $n^{th}$ order term 
will go like $g^n\beta^n$ (subleading corrections in $\beta$ are present). 
Here, in first order, we have a whole series in $g$ times $\beta$. Thus, 
if we remember that the {\it exact} ground state $E_0(g)$ is the only 
state to contribute at large $\beta$:
\begin{equation}
Z(\beta) \stackrel{\beta \to \infty}{\longrightarrow}
e^{-\beta E_0(g)}=e^{-\frac{\beta\hbar\omega}{2}}
\left\{ 1- \beta\left[ E_0(g)-\frac{\hbar\omega}{2}\right]
+ O(\beta^2) \right\},
\end{equation}
we see that we obtain a nonperturbative estimate for $E_0(g)$, while 
perturbation theory would only give us an order $g$ correction.

In order to compute correlation functions, one might introduce an 
external current, $j(\tau)$, coupled to the variable, $x(\tau)$, and 
use the knowledge of the semiclassical propagator to integrate the linear
plus quadratic piece of the action to derive a functional of $j(\tau)$, to 
be used in the expansion in the standard way \cite{amit}.

%%%%%%%%%%%%%%%%%%%%%%%%%%%%%%%%%%%%%%%%%%%%%%%%%%%%%%%%%%%%%%%%%%%%%%%%%

\section{Conclusions and outlook}

The methods developed in the preceding sections allow us to deal with 
problems in one-dimensional Quantum Statistical Mechanics by means of a 
semiclassical expansion which is fully calculable and yields 
nonperturbative results.

Problems in higher-dimensional Quantum Statistical Mechanics with 
potentials that have spherical symmetry may be reduced, by means of 
partial-wave decompositions, to one-dimensional problems wherein our 
methods could be applied.

Furthermore, in finite temperature Field Theory, where the point $x_0$ 
of our previous treatment is to be replaced with the (static) field 
configuration, $\phi(\vec{x})$, we already see the need to consider 
extrema which must depend on $\tau$, the Euclidean time, as we lower 
the temperature. Catastrophes will certainly occur, forcing us to 
rethink commonly used approximations that use constant field 
configurations as backgrounds to derive effective potentials. Here, 
again, under very special circumstances, one might hope that the 
methods developed so far may shed some light into the intrincacies of 
nonperturbative phenomena.

\acknowledgments

This work was partially supported by CNPq, FUJB/UFRJ and CLAF. It is 
a pleasure to thank Eduardo Fraga and Sergio Jor\'as for help with the 
manuscript, as well as for comments and suggestions. Special thanks to
the organizers of the La Plata meeting for their kind hospitality.

%%%%%%%%%%%%%%%%%%%%%%%%%%%%%%%%%%%%%%%%%%%%%%%%%%%%%%%%%%%%%%%%%%%%%%% 

\newpage 
  
\noindent 
\underline{\bf Figure Captions}: 
  
\vspace{5mm} 
 
\noindent 
{\bf Figure 1}: (a) Classical paths at $x_m$ for $\beta<\pi/\hbar\omega_m$ 
(1) and $\beta>\pi/\hbar\omega_m$ (2 and 3); (b) Sketch of how the  
extrema change along the unstable direction in function space. 
 
\vspace{5mm} 
 
\noindent 
{\bf Figure 2}: (a) Classical paths at $x_m$ for $\beta<\beta_{\rm c}$ (1),  
$\beta=\beta_{\rm c}$ (1 and 2\&3) and $\beta>\beta_{\rm c}$  
(1,2 and 3), $\pi/\hbar\omega_m<\beta_{\rm c}<2\pi/\hbar\omega_m$; 
(b) Sketch of how the extrema change along the unstable direction  
in function space. 
 
\vspace{5mm} 
 
\noindent  
{\bf Figure 3}: Bifurcation set for the cusp; pattern of extrema shown  
schematically. 
 
\vspace{5mm} 
 
\noindent 
{\bf Figure 4}: Evolution of the classical action as $\beta$ changes.  
(a) $0\le\beta<\pi/\hbar\omega_m$; 
(b) $\pi/\hbar\omega_m<\beta<2\pi/\hbar\omega_m$; 
(c) $2\pi/\hbar\omega_m<\beta<3\pi/\hbar\omega_m$. 
 
\vspace{5mm} 
 
\noindent 
{\bf Figure 5}: Partition of control space into $p$-solution regions  
$(p=1,3,5,7,\ldots)$. 
 
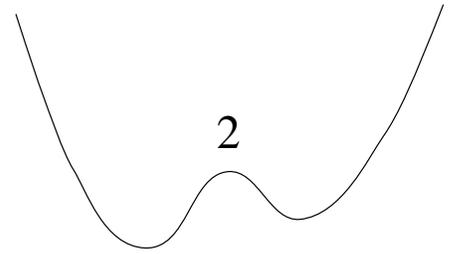
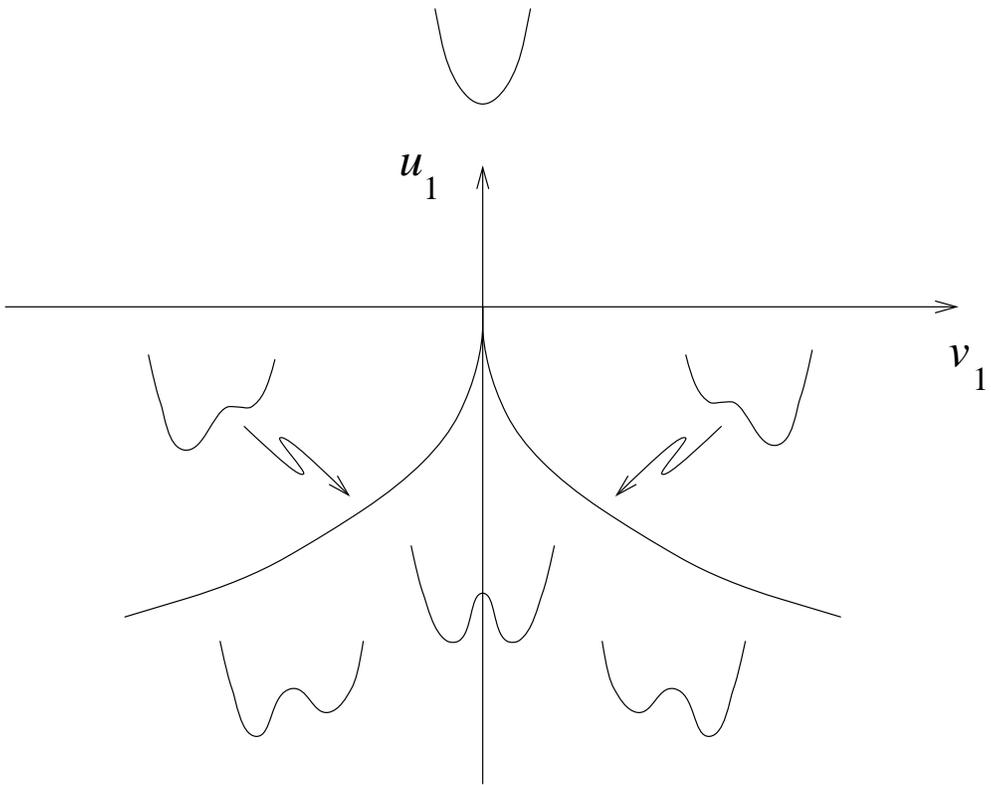
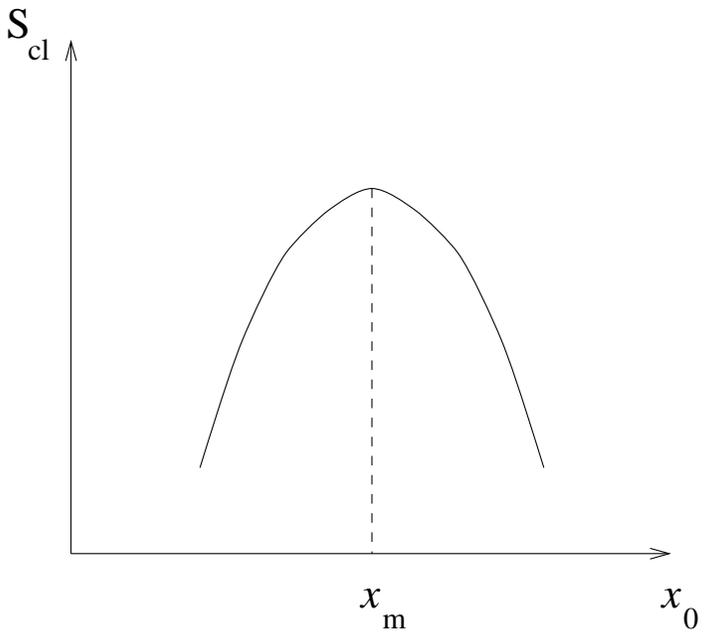
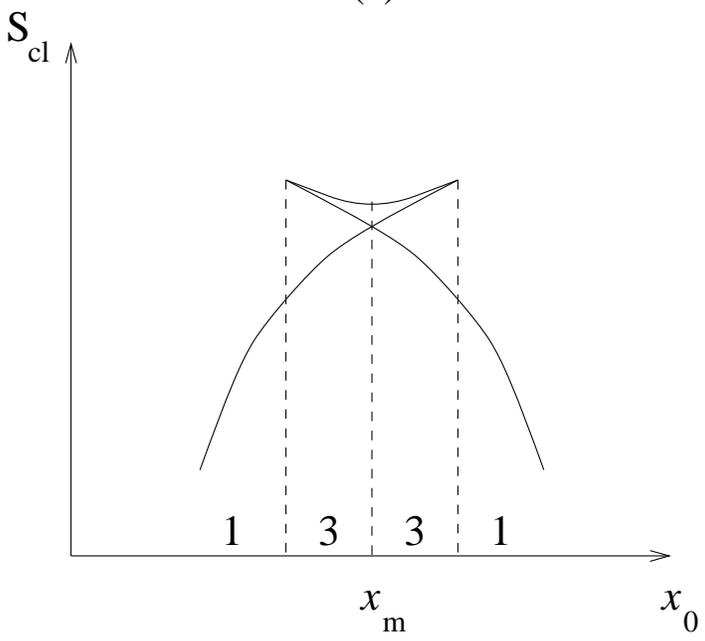
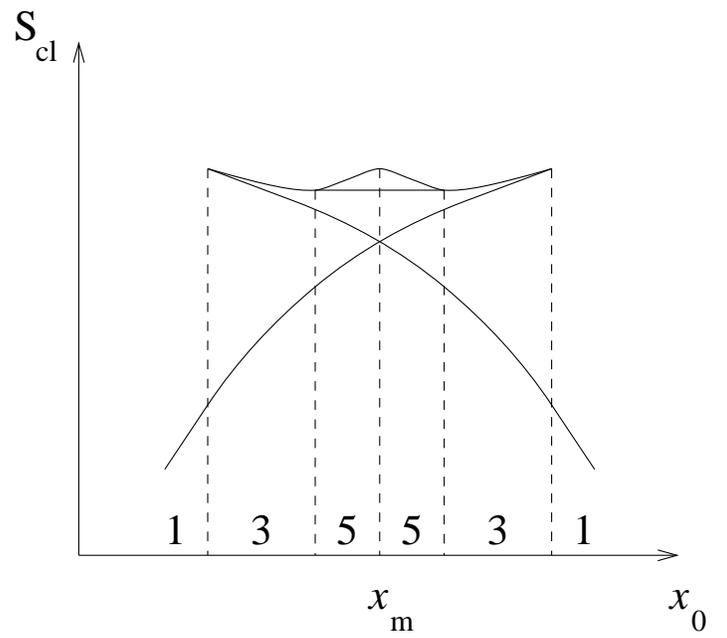
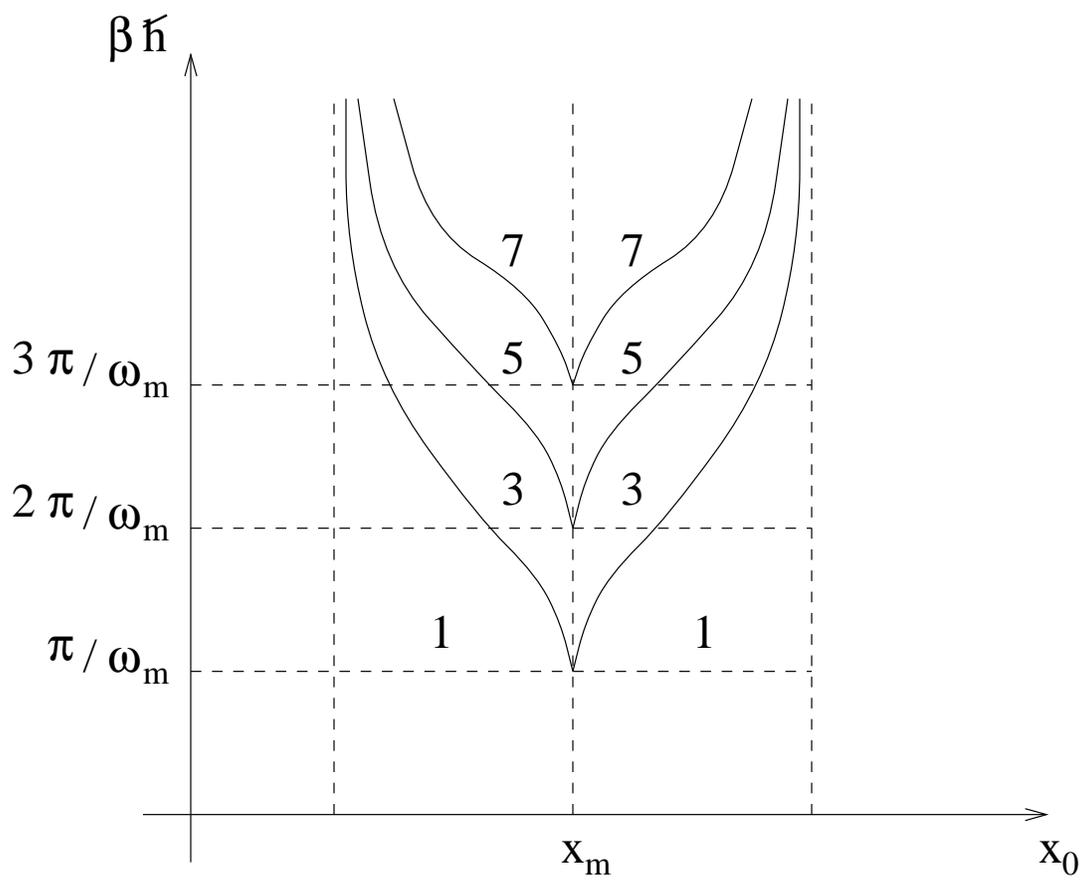
\end{document}